\documentclass[twocolumn,showpacs,preprintnumbers,amsmath,amssymb]{revtex4}

\usepackage{graphicx}
\usepackage{epsfig}

\usepackage{longtable}
\usepackage{longtable}

\begin{document}

\title{ Candidates for Long Lived High-K Ground States in Superheavy Nuclei }

\author{P.~Jachimowicz}
 \affiliation{Institute of Physics,
University of Zielona G\'{o}ra, Szafrana 4a, 65516 Zielona
G\'{o}ra, Poland}

\author{M.~Kowal} \email{m.kowal@fuw.edu.pl}
\affiliation{National Centre for Nuclear Research, Ho\.za 69,
PL-00-681 Warsaw, Poland}

\author{J.~Skalski}
\affiliation{National Centre for Nuclear Research, Ho\.za 69,
PL-00-681 Warsaw, Poland}

\date{\today}

\begin{abstract}
 On the basis of systematic calculations for 1364 heavy and superheavy nuclei, including odd-systems,
  we have found a few candidates for high-K ground states in superheavy nuclei.
 The macroscopic-microscopic model based on the deformed Woods-Saxon
 single particle potential which we use offers
 a reasonable description of SH systems, including known: nuclear masses, $Q_{\alpha}$-values, fission barriers,
 ground state deformations, super- and hyper-deformed minima in the heaviest nuclei.
 Exceptionally untypical high-K intruder contents of the g.s. found for some
 nuclei accompanied by a sizable excitation of the parent configuration in
 daughter suggest a dramatic hindrance of the $\alpha$-decay.
 Multidimensional hyper-cube configuration -
 constrained calculations of the Potential Energy Surfaces (PES's) for one
 especially promising candidate, $^{272}$ Mt, shows a
 $\backsimeq$ 6 MeV increase in the fission barrier above the configuration-
 unconstrained barrier.
 There is a possibility, that one such high-K ground- or low-lying state may
 be the longest lived superheavy isotope.
\end{abstract}


\maketitle


  Two current methods of making superheavy elements in the
 laboratory: “cold” \cite{GSI,GSI2} and “hot” \cite{O1} fusion reactions,
 seem to reach their limits.
 In the first one, (i) evaporation residues (ERs) produced on Pb or Bi targets
 are far from the predicted “island of stability”, (ii) ER cross sections
 drop steeply with the mass of a projectile.
 In the latter, with the $^{48}$Ca projectile impinging on various actinide
 targets, ERs cross sections are roughly constant. However, once the element
 Z=118 has been synthesized \cite{O2}, attempts to go beyond hit two obstacles:
 (i) a difficulty or impossibility of making targets from Es and heavier
 actinides, (ii) reactions with heavier projectiles like: $^{50}$Ti, $^{54}$Cr,
  $^{58}$Fe, and $^{64}$Ni, did not produce any ERs up to now.

 On the other hand, not all superheavy (SH) isotopes $Z\leq 118$ have been
 produced yet. It might be, that among them hides some surprisingly long-lived
 one, either in its ground- or excited isomeric state. It is even not excluded
 that such a long-lived SH state was already produced, but remained undetected.
 Since detection procedures used at present are adjusted to short-time
 coincidences, a species living tens of minutes would be very likely missed
 in the background. Therefore, while pondering upon possible new reactions
 leading towards the island of stability, it may be worthwhile
  to search for a long-lived exotic SH configurations.
 Obvious candidates are high-K isomers or ground-states, for which increased
 stability is expected due to some specific hindrance mechanisms.
 The present letter provides predictions and arguments for such long-lived SH
  states.

The existence of isomeric states is rather well establish in nuclear structure physics.
 The structure of expected long-lived multiquasiparticle high-spin isomers in
 some even-even SH nuclei was analyzed e.g. in \cite{Xu}. In particular,
 the assignments of 9$^{-}$ or 10$^{-}$  \cite{Hof}
 two quasineutron configurations for the $6.0 ^{+8.2}_{-2.2}$ms isomer in
 $^{270}$Ds (the heaviest isomer known) was supported \cite{Xu}.
 Let us stress that the half-life of this isomer is much longer
 than that of the ground state ($100 ^{+140}_{-40}$ $\mu$ s). The same holds
 for the 8$^{+}$ isomer in $^{256}$Es, with the half-life of 7.6h -
 significantly longer than 25 min of the g.s. Another interesting example
 is a 16$^{+}$ or 14$^{+}$ state in $^{254}$No, with a half-life 184 $\mu$s,
  at 2.93 MeV above g.s \cite{Her,Tan,Kon,Hes,Cla}. In \cite{Liu1},
  the four-qp isomers around $^{254}$No were postulated. All examples
  described above relate to prolate equilibria. A possibility of high-K
  isomers  at the superdeformed oblate minima in SH nuclei was indicated
 in \cite{SDO}.

 The electromagnetic stability of an isomeric state is difficult to predict
 as it depends on fine details of the s.p. structure.
  Therefore, we concentrate here on high-K ground states or very-low-lying
  configurations in odd-odd nuclei.
  Excited configurations are also of interest, as some detected $\alpha$
 transitions,
 and even whole portions of $\alpha$ decay chains, may actually connect not
 g.s., but excited states of similar structure.
  The hindrance of the s.f. in odd-odd relative to even-even isotopes by
  several orders of magnitude is well established in heavy nuclei.
  It is also known from experimental studies that $\alpha$-decays
  accompanied by a change of the parent configuration are hindered with respect
  to configuration-preserving transitions. It is understood as a decrease in
  the probability of the $\alpha$-particle formation when different
 parent and daughter configurations are involved.
   Alpha-decay rate shows also the exponential dependence on the
   barrier between $\alpha$ particle and the daughter nucleus.
   When a parent configuration has some excitation energy
 in daughter, this barrier increases and an effective $Q_{\alpha}$ value
   decreases by this energy.
  If the configuration-changing decays had been hindered completely, the
  $\alpha$-decay rate would have been given by the reduced $Q_{\alpha}$
  values.
  In reality, a degree of the configuration hindrance is surely
  configuration-dependent.

 In order to predict possible exotic configurations one has to have a
  reliable model to find ground states in odd and odd-odd nuclei.
 We used the macroscopic-microscopic model based on a deformed Woods-Saxon
 potential \cite{WS} and the Yukawa plus exponential macroscopic energy \cite{KN}
 Recently, within this approach (with parameters adjusted
 to heavy nuclei \cite{WSpar}), it was possible to reasonably reproduce data on
 first \cite{Kow},  second \cite{kowskal,IIbarriers} and third \cite{IIIbarriers1,IIIbarriers2} fission barriers and
 systematically predict ground states and saddle-points in even-even
 superheavy nuclei up to $Z=126$ \cite{archive}.

  For systems with odd proton or neutron (or both), we used a standard
  blocking method.
 The ground states were found by minimizing over configurations (blocking
 particles on levels from the 10-th below to 10-th above the Fermi level)
 and deformations. For nuclear ground states it was possible to confine
 analysis to axially-symmetric shapes.
 In the present study, four mass- and axially-symmetric
 deformations: $\beta_{20}, \beta_{40}, \beta_{60} , \beta_{80}$
  were used - see \cite{Jach2014}.
 Minimization is performed by the gradient method and
 on the mesh of deformations, see \cite{Jach2014}.
 Both sets of results are consistent.
 As an additional check, we performed another calculations of nuclear masses
 within the quasiparticle method.
 These calculations are much simpler and we were
 able to perform a seven-dimensional minimization over axially-symmetric
 deformations:$\beta_{20}, \beta_{30}, \beta_{40}, \beta_{50}, \beta_{60}, \beta_{70}$ and $\beta_{80}$.

 A simplest extension of the WS model to odd nuclei required three new
  constants which may be interpreted
 as the mean pairing energies for even-odd, odd-even and odd-odd nuclei
 \cite{Jach2014}.
 They were fixed by a fit to the masses with $Z\geq82$ and $N>126$ via
 minimizing the rms deviation in particular groups of nuclei what is
 rather standard procedure \cite{mol95,mol97}. Masses of nuclei were taken from \cite{Wapstra2003}.
 The obtained rms deviation in masses for 252 nuclei was about
 400 keV with blocking and around 300 keV with the q.p. method. Similar rms
 errors resulted for 204 $Q_{\alpha}$ values. Then, for 88 measured
 $Q_{\alpha}$ values in SH nuclei, the quantities outside the region of the
 fit, we obtained the rms deviation of about 250 keV, similar for two
 treatments of pairing.

\begin{figure}[h]
\hspace{-1.8cm}
\includegraphics[scale=0.4]{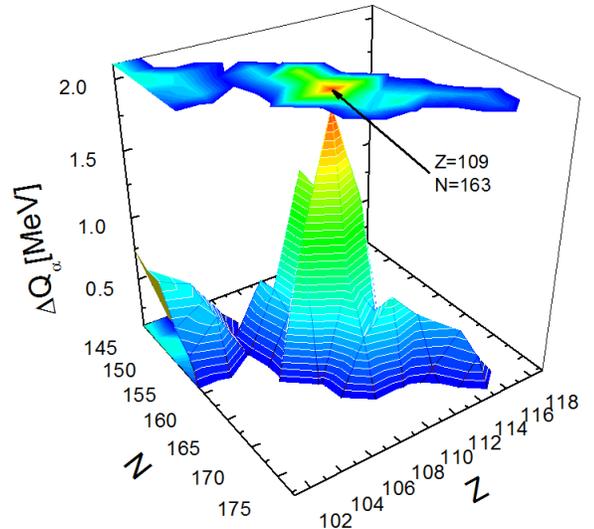}
\vspace{-1.0cm}
\caption{Excitation of the parent g.s. configuration in the
 daughter nucleus  $\Delta Q_{\alpha}$ in isotopes with various Z and N.}\label{3D}
\end{figure}

 We also calculated apparent $Q_{\alpha}$ values taking
  the parent g.s. configuration as the final state in daughter.
   Such a value is smaller than the true $Q_{\alpha}$ by the
   excitation energy of the parent g.s. configuration, $\Delta Q_{\alpha}$,
   in the daughter.
   Global results are collected and shown in Fig. \ref{3D}.
   Particularly large, about 2 MeV, excitation of the parent configuration
  occurs in the daughter of $^{272}$Mt; the excitation of
  about 1.5 MeV occurs for $^{256}$Lr  thus energies of these configuration-preserving transitions
  are reduced especially for particle numbers corresponding to one particle
  above a closed subshell.

  Predicted high-K g.s.configurations in odd and odd-odd SH nuclei whose
 excitation energy in the daughter is larger than 0.5 MeV are shown in
 Table \ref{t1}. They result from an interplay between high-$j$ orbital
 positions and energy minimization.
  A particular situation occurs at deformed subshells $Z=102,108$ and
  $N=152, 162$. The levels just above gaps include intruder
 $\Omega^{\pi}=$ 9/2$^+$ and 11/2$^+$ protons and 11/2$^-$ and 13/2$^-$
 neutrons. It turns out that high-K ground states involving two intruders occur
  in $^{262}$Db and $^{272}$Mt.
  In Table \ref{t1} we shown the g.s. configurations whose excitation in the daughter is larger than 0.5 MeV.
  It is likely that they have elongated $\alpha$-decay half-lives.

\begin{table}[tbh]
\caption{Predicted high-K ground states in SH nuclei whose excitation
 in daughter is larger than 0.5 MeV.}
\label{t1}
\begin{tabular}{llllll|lllllll}
 $Z$ & $N$ & $\Omega^p$ & $\Omega^n$ & $K^{\pi}$ & $\Delta Q_{\alpha}$ & $Z$ & $N$ & $\Omega^p$ & $\Omega^n$ & $K^{\pi}$ & $\Delta Q_{\alpha}$\\
\hline
111    &    169    &    9/2$^-$    & 5/2$^+$     &    7$^-$       &    0.74    \\
111    &    163    &    3/2$^-$    & 13/2$^-$    &    8$^+$       &    1.31    & 108    &    163    &               & 13/2$^-$    &    13/2$^-$    &    1.00    \\
111    &    161    &    3/2$^-$    & 7/2$^+$     &    5$^-$     &    0.52    & 107    &    163    &    5/2$^-$    & 13/2$^-$    &    9$^+$       &    1.17    \\
110    &    163    &               & 13/2$^-$    &    13/2$^-$    &    0.97    & 107    &    157    &    5/2$^-$    & 11/2$^-$    &    8$^+$       &    0.57    \\
109    &    169    &    11/2$^+$    & 9/2$^+$    &    10$^+$      &    0.51    & 106    &    163    &               & 13/2$^-$    &    13/2$^-$    &    0.96    \\
109    &    167    &    11/2$^+$    & 5/2$^+$    &    8$^+$       &    0.71    & 105    &    153    &    9/2$^+$    & 1/2$^+$     &    5$^+$       &    0.97    \\
109    &    166    &    11/2$^+$    &            &    11/2$^+$    &    0.88    & 103    &    157    &    7/2$^-$    & 11/2$^-$    &    9$^+$       &    0.52    \\
109    &    165    &    11/2$^+$    & 3/2$^+$    &    7$^+$       &    1.38    & 103    &    154    &    7/2$^-$    &             &    7/2$^-$     &    0.54    \\
109    &    164    &    11/2$^+$    &            &    11/2$^+$    &    1.13    & 103    &    153    &    7/2$^-$    & 1/2$^+$     &    4$^-$       &    1.45    \\
109    &    163    &    11/2$^+$    & 13/2$^-$   &    12$^-$      &    1.99    & 103    &    151    &    7/2$^-$    & 9/2$^-$     &    8$^+$       &    0.58    \\
109    &    162    &    11/2$^+$    &            &    11/2$^+$    &    1.27    & 103    &    149    &    7/2$^-$    & 7/2$^+$     &    7$^-$       &    0.63    \\
109    &    161    &    11/2$^+$    & 9/2$^+$    &    10$^+$      &    1.32    & 101    &    151    &    1/2$^-$    & 9/2$^-$     &    5$^+$       &    0.68    \\
109    &    160    &    11/2$^+$    &            &    11/2$^+$    &    1.37    & 101    &    149    &    1/2$^-$    & 7/2$^+$     &    4$^-$       &    0.92    \\
109    &    159    &    11/2$^+$    & 9/2$^+$    &    10$^+$      &    1.56    \\
109    &    158    &    11/2$^+$    &            &    11/2$^+$    &    1.39    \\
109    &    157    &    11/2$^+$    & 3/2$^+$    &    7$^+$       &    1.41    \\
\hline
\end{tabular}
\end{table}

\begin{figure}[h]
\includegraphics[width=\columnwidth]{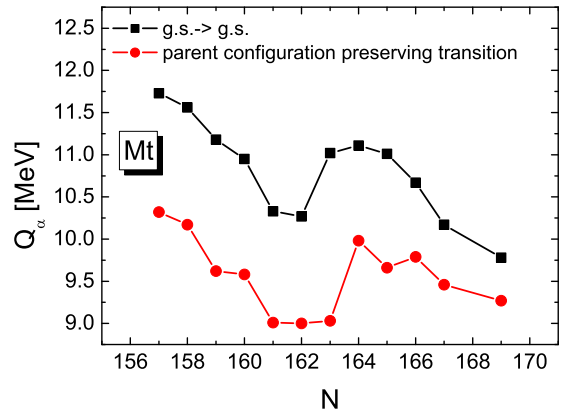}
\includegraphics[width=\columnwidth]{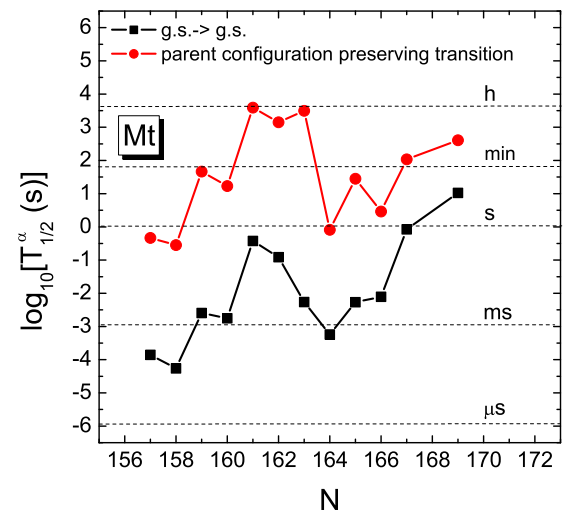}
\caption{{\label{fig3}
 $Q_{\alpha}$ values calculated with blocking following from the WS model
  and the increase in $\alpha-$ half-life induced by $\Delta Q_{\alpha}$
  \cite{Roy}, assuming a complete decay hindrance to
  different configurations.}
\label{QalphaTalpha}  }
\end{figure}

  In Fig. \ref{QalphaTalpha}, a decrease in energy release,
 $\Delta Q_{\alpha}$, and $\log_{10} T_{\alpha}$ calculated
  according to \cite{Roy} for g.s.$\rightarrow$ g.s.  and
 configuration-preserving transitions are shown for
  various Meitnerium isotopes. The latter half-life would correspond
 to an absolute hindrance of configuration-changing decays.
 For our favorite case, $^{272}$ Mt, such hindrance reaches six orders of
 magnitude. This takes us from the life-time of milliseconds to hours.

 If one assumes that the configuration hindrance is not absolute but, for example,
 it amounts to 100 per unpaired nucleon state with the same $S_z$ value ($\Sigma$ in the Nilsson notation) as in the parent,
 than one obtains the hindrance factors for various final states in the daughter $^{268}$Bh shown in Table \ref{t2}.
 The orbitals forming the chosen configurations have the expectation value of $\Sigma$ very close to 1, as both blocked orbitals in the g.s. of $^{272}$Mt.
 Their deformations, obtained by energy minimization, are similar as for the ground state of $^{268}$Bh.
 The hindrance from the $Q_{\alpha}$ reduction, HF$(\Delta Q_{\alpha})$, is included; this means the total hindrance is: $10^{4} \times $ HF$(\Delta Q_{\alpha})$.
 As may be seen, the lowest hindrance corresponds to the configuration nearly degenerate with the ground state.
 Thus, the result of such an estimate is implied by the assumed configuration hindrance per orbital.
\begin{table}
\caption { Specification of final states in $^{268}$Bh for protons ($\pi$) and neutrons ($\nu$) with appropriate expectation values   $<\Sigma^{\pi(\nu)}>$.
The corresponding $Q_{\alpha}$ energies in [MeV] and life-times $\log_{10} T_{\alpha}$ in [s] of $^{272}$Mt are shown.
First line concerns ground state to ground state transition.}
\label{t2}
\begin{ruledtabular}
\begin{tabular}{ccccccc}
$\pi$ - state  & $<\Sigma^{\pi}>$  & $\nu$ - state & $<\Sigma^{\nu}>$&  $Q_{\alpha}$   & $\log_{10} T_{\alpha}$  \\
\noalign{\smallskip}\hline\noalign{\smallskip}

5/2$^-$ [512] & 0.93  & 9/2$^+$  [615]  & -0.86  & 11.02 & -2.26   \\
5/2$^-$ [512] & 0.93  & 7/2$^+$  [613]  & 0.88   & 10.93 &  1.97   \\
5/2$^-$ [512] & 0.93  & 11/2$^-$ [725]  & 0.92   & 10.64 &  2.70   \\
9/2$^+$ [624] & 0.91  & 7/2$^+$  [613]  & 0.88   & 10.31 &  3.60   \\
9/2$^+$ [624] & 0.91  & 11/2$^-$ [725]  & 0.92   & 10.12 &  4.14   \\
\noalign{\smallskip} \noalign{\smallskip}
\end{tabular}
\end{ruledtabular}
\end{table}

 The measured $\log_{10}T_{\alpha}$ values for Mt isotopes are close
 to the lower line in Fig. \ref{QalphaTalpha}, except for $N=161$ (5 ms \cite{Gupta05}) which
 is below, and $N=165$ (0.44 s \cite{Gupta05}) which is above it. Note, that the
 supposedly isomeric half-life, nearly identical with the lower theoretical
  point for $N=161$ was reported \cite{Morita04}.
  We claim that half-lives corresponding to the upper
  theoretical curve have not been detected yet, maybe, except for $N=165$.

  Let us stress that the whole argument is based on both the
 presence of deformed semi-magic shell Z=108 and N=162 and the
 position of high-$\Omega$ intruder orbitals just above that shell.
 The former is supported by the experimental $Q_{\alpha}$ values and
 predictions by many models. The latter is more model-dependent, but
 the same or nearly the same intruder positions are predicted by
 the FRLDM \cite{mol97} and the SLy4 HFB \cite{CNH99}.

 In order to estimate fission rate for a high-K ground state one
 can start with the known half-lives of the neighbours, the calculated barriers and half-lives in even-even nuclei \cite{SmolSkalSob}, and then
 consider the effect of blocking on the fission barrier and non-adiabatic effects in barrier tunneling.
  The experimental half-lives in $N=163$ isotones are 11 h in $^{266}$Lr (s.f.), 1.3 h in $^{267}$Rf (s.f.), 30.8 h in $^{268}$Db (s.f. + EC),
  2 m in $^{269}$Sg ($\alpha$),  3.8 m in $^{270}$Bh ($\alpha$) and 4 s in $^{271}$Hs ($\alpha$).
  Since the g.s. shell effect in $^{270}$Hs is necessarily related to a
  local barrier maximum \cite{Kow}, its s.f. half-live is in the range of hours \cite{SmolSkalSob}.
  This suggests a spontaneous fission half-life of $^{272}$Mt, the odd-odd nucleus, not less than that.

\begin{figure}[h]

\vspace{-7mm}

\begin{minipage}[l]{0mm}

\centerline{\includegraphics[scale=1.0]{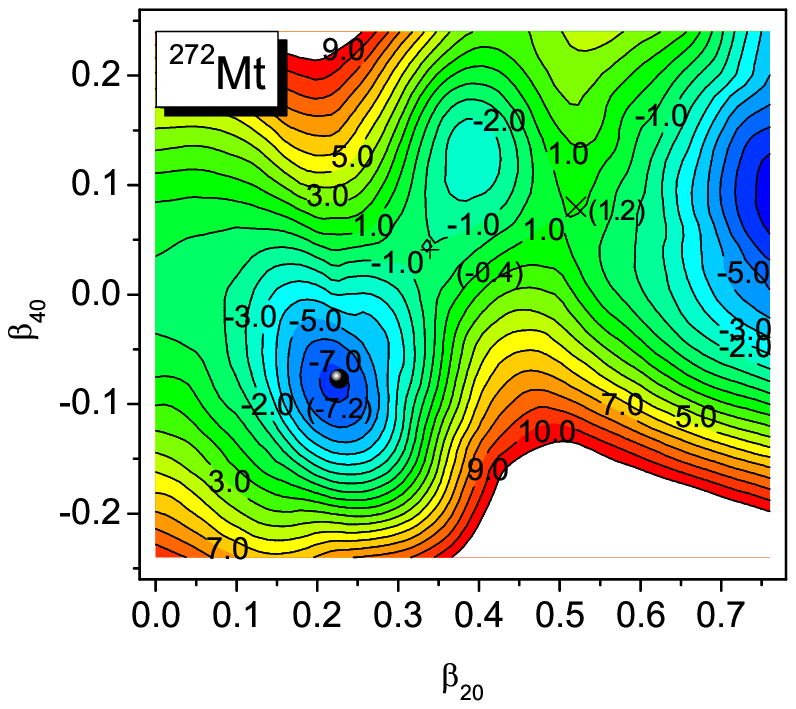}}
\centerline{\includegraphics[scale=1.0]{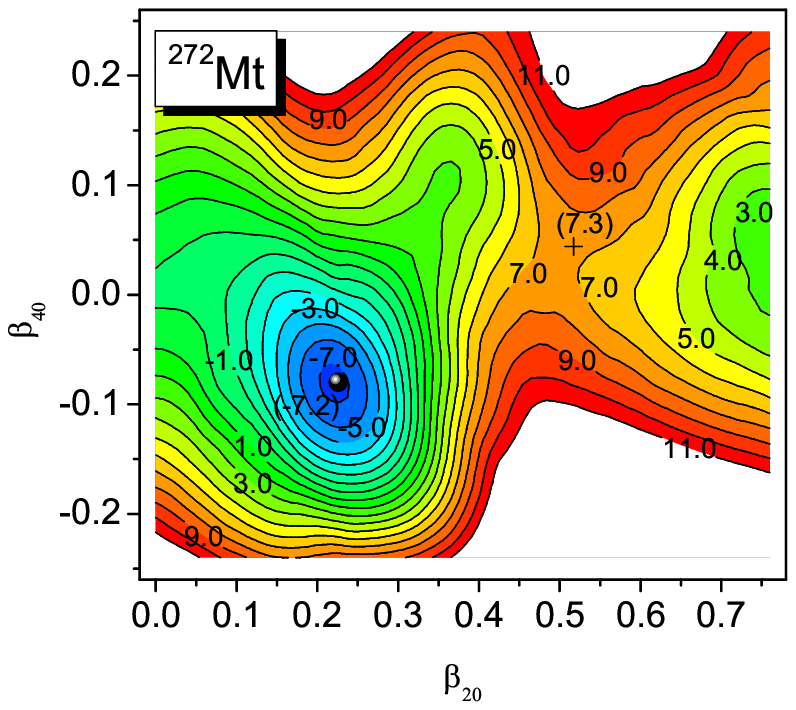}}
\end{minipage}
\vspace{-2mm} \caption{{\protect Energy surfaces: $E-E_{mac}(sphere)$  for $^{272}$ Mt heavier system in two extreme situations i.e.
adiabatic with minimization over possible configurations (top panel) and non-adiabatic configuration-constrained calculations  \label{Mt} }}
\end{figure}

The calculated fission barrier {\it minimized over configurations} in
$^{272}$Mt amounts to 6.3 MeV, nearly the same as in in $^{270}$Hs.
 The 5D calculation, including quadrupole nonaxility in addition to four
  axially- and reflection-symmetric deformations, was done on the hypercube
 and saddles were obtained by the immersion method.
 In the near-by even-even isotopes, the nonaxiality in the first barrier was
  shown not important \cite{Kow}.
 To see the configuration blocking effect,
 two additional 4D saddle point calculations were performed for axially
 symmetric deformations in two extreme scenarios: diabatic (with the blocked
 g.s. configuration) and adiabatic one (minimized over configurations).
 Also here, saddle points were determined by the immersion method.
 In Fig. \ref{Mt} are shown $\beta_{20}$, $\beta_{40}$ energy maps, obtained
  by minimization over the remaining deformations.  One can see a rather
 dramatic effect of  keeping the g.s. configuration: the diabatic saddle point
 (bottom panel of Fig.  \ref{Mt}) lies about 6 MeV higher than the
 adiabatic one (top panel of Fig. \ref{Mt}).
 Moreover, there is no second (local) minimum on the diabatic map.
 We have also checked that the reflection-asymmetry effect on the
 diabatic/adiabatic saddle is small.

  It is known that the blocking procedure often causes an excessive reduction
  of the pairing gap in systems with odd particle number. One device to avoid
  an excessive even-odd staggering in nuclear binding was to assume
  stronger (typically by $\sim$ 5\%) pairing interaction for
  odd-particle-number systems, see e.g. \cite{Gor0}.
  To estimate the effect of a stronger pairing on s.f. barriers
 we have calculated barriers with pairing strengths
 increased by 10\% and 15\%. We obtained still large diabatic
   fission barriers of 11.8 MeV and 10.3 MeV, respectively.

  Obviously, one cannot expect a strict conservation of configuration
  during the tunneling. It is unavoidable, however, that
 any configuration change, either related to nonaxiality or to the collective
 rotation admixture, must induce some increase in the action integral,
 relative to the neighbouring even-even nucleus.
 Therefore, the fission half-life of the high-K ground state of $^{272}$Mt in the range of hours seems to be a safe prediction.

 When it comes to $\beta$ decay,
 it may be added that the predicted Q$_{\beta}$ value of 4.2 MeV in $^{272}$Mt means that its $\beta$ half life is in the range of minutes \cite{MOL}.

%
To summarize:

(i) The macro-micro WS model, which has been shown to give reasonable
 predictions in the SH region, suggests high-K configurations as the ground
 states in a number of odd and odd-odd nuclei.

 (ii) An $\alpha$-decay hindrance of such a configuration is expected
    when the same configuration in daughter has a sizable excitation.
   If configurantion-changing transitions had been strictly forbidden,
   the hindrance would have been determined by this excitation
 $\Delta Q_{\alpha}$.

(iii) A particular situation occurs above double closed subshells:
 $N=162$ and $Z=108$ where two intruder orbitals: neutron $13/2^-$ from
 $j_{15/2}$ and proton $11/2^+$ from $i_{13/2}$ spherical subshells are
 predicted.
 These orbitals combine to the $12^-$ g.s. in $Z=109$, $N=163$, whose
 configuration lies $\sim$2 MeV above the g.s. the daughter.
 This would imply a six order of magnitude increase in $\alpha$ half-life.

 (iv) The double subshell gap at $N=162$ and $Z=108$ is consistent with
     experimental $Q_{\alpha}$  values and predicted by many models.
  The position of neutron $13/2^-$ and proton $11/2^+$ orbitals above this gap
  is also common to many models, in particular, the predicted
  g.s. in  $^{272}$Mt has exactly the same structure in \cite{mol97};
  the same configuration is predicted at $\sim$0.4 MeV excitation within the
  SLy4 HFB model \cite{CNH99}, which would make it also a long-lived one.

(v) The calculated configuration-preserving fission barrier in $^{272}$Mt
is by 6 MeV higher than the one minimized over configurations.
    Even if configuration is not completely conserved, a substantial increase
 in fission half-life is expected.

 (vi) There are other orbitals which may produce long-lived configurations, in
 particular intruder neutron $11/2^-$ and proton $9/2^+$  above
 $N=152$, $Z=102$. Although more model-dependent, there is a possibility, that
 one such high-K ground- or low-excited state may be the longest lived
 superheavy nucleus.

 As already mentioned, one cannot exclude that such long-lived superheavy
 configurations were already produced before, but setup and electronics
 dedicated to the milliseconds measurements could not detect decays of objects
 living much longer. This intriguing possibility should be check at first.

M.K. and J.S. were co-financed by the National Science Centre under Contract
No. UM0-2013/08/M/ST2/0025 (LEA-COPIGAL). One of the authors
(P.J.) was cofinanced by Ministry of Science and Higher Education:
„Iuventus Plus” grant Nr IP2014 016073.
This research was also supported by an allocation
of advanced computing resources provided by the Swierk Computing Centre (CIS)
 at the National Centre for Nuclear Research (NCBJ) (http://www.cis.gov.pl).

\end{document}